\documentclass[12pt]{article}

\newcommand{\beq}{\begin{eqnarray}}
\newcommand{\eeq}{\end{eqnarray}}
\newcommand{\beqs}{\begin{eqnarray*}}
\newcommand{\eeqs}{\end{eqnarray*}}

\usepackage{graphics,epsfig}
\usepackage{latexsym}
\usepackage{amsthm}
\usepackage{amssymb,amsfonts,amsmath}
\usepackage{pdflscape}

\begin{document}

\title{Disaggregation of Small, Cohesive Rubble Pile Asteroids due to YORP}
\author{D.J. Scheeres\\Department of Aerospace Engineering Sciences \\ The University of Colorado\\ Boulder, Colorado 80309 \\ scheeres@colorado.edu}
\date{}
\maketitle

\begin{abstract}
The implication of small amounts of cohesion within relatively small rubble pile asteroids is investigated with regard to their evolution under the persistent presence of the YORP effect. We find that below a characteristic size, which is a function of cohesive strength, density and other properties, rubble pile asteroids can enter a ``disaggregation phase'' in which they are subject to repeated fissions after which the formation of a stabilizing binary system is not possible. Once this threshold is passed rubble pile asteroids may be disaggregated into their constituent components within a finite time span. These constituent components will have their own spin limits -- albeit potentially at a much higher spin rate due to the greater strength of a monolithic body. The implications of this prediction are discussed and include modification of size distributions, prevalence of monolithic bodies among meteoroids and the lifetime of small rubble pile bodies in the solar system. The theory is then used to place constraints on the strength of binary asteroids characterized as a function of their type. 
\end{abstract}

\section{Introduction}

That asteroids are rubble pile bodies is no longer a controversial theory, with the real questions migrating to what the relative mix of rubble pile to monolithic asteroids may be. There are many specific observations that support the idea of rubble pile asteroids, including the presence of a spin limit for larger asteroids, the uniformly high porosity observed in asteroids as compared to their expected meteorite matches, and the direct observations and measurements of asteroids from spacecraft, such as occurred for asteroids Mathilde, Itokawa and Eros. The evidence and implications of these observations are based on decades of observations and has been summarized recently in \cite{scheeres_AIV}. 


Not as commonly discussed is the connection between the rubble pile structure of asteroids and the size distribution of their population. That these are inextricably linked is clear, as the rubble pile structure of asteroids is created from catastrophic disruptions which also create asteroid families and forms the fundamental process that shapes the size distribution. The relevant work in the area of impacts is summarized in \cite{bottke_AIV}. 

The distinction between rubble pile structure and population is important, but not necessarily linked for larger asteroids. However, smaller asteroids -- specifically those less than $\sim$ 10 km in size -- are physically modified by the non-gravitational YORP effect. One of the implications of the YORP effect is that the spin rates of smaller asteroids are changed over timescales that are short relative to their collisional lifetime in the main belt, or their dynamical lifetime in the Near Earth Asteroids (NEA) population \cite{gladman_2000}.  When their spin rates become large enough YORP causes rubble pile asteroids to deform (hence the genesis of the spin deformation limit) and to undergo fission processes which can cause rubble pile bodies to separate into multiple components and, through dynamical and physical evolution, create binary or multiple component asteroid systems. On average these fission processes will cause a decrease in any particular asteroid's size, through loss of one of the components in creation of an asteroid pair or through the direct shedding of material that escapes from the parent body. The timescale over which these fission processes will reduce the size of an asteroid can be long, and not solely tied to the YORP effect. In particular, binary asteroids can settle into long-term stable configurations that balance tides against the Binary YORP effect \cite{jacobson_BYORP_equilibrium} or for doubly synchronous system balance non-gravitational forces in a stable equilibrium \cite{golubov_DSBYORP}. Even if an asteroid does not settle into one of these configurations, the tidal timescales of a multicomponent system can be very long \cite{jacobson_mcmahon_wide} -- essentially conserving the asteroid's mass over long time periods. Despite this, such asteroids undergo a slow erosion, losing the components of their rubble pile structure either as individual pieces, or perhaps more commonly by separating into components that are themselves rubble piles. Such processes have been shown to be consistent with the current structure of the asteroid size population, e.g. \cite{jacobson_NEA_pop}. 

This gradual erosion also raises an interesting question regarding the relative size distribution of the asteroid population and the size distribution of strong components within a rubble pile body. Data on the latter are limited, with the most precise information coming from observations of Eros and Itokawa. Their components, i.e., competent boulders and grains, would in general have different physical properties than rubble pile asteroids, specifically they would be able to spin faster \cite{holsapple_spinlimits} and would have larger densities. They would also not be subject to the same fission processes that rubble pile asteroids are subject to, of course. The size that such components could reach is a question of some interest, and may be visible in the population as fast spinning bodies that must have strength greater than that predicted for rubble pile bodies. These bodies may already be observed as the fast rotators in the asteroid population (first reported in \cite{pravec2002large} and more recently seen in \cite{chang2014new}). However, as has been noted previously \cite{holsapple_spinlimits, sanchez_MAPS}, for smaller asteroids the spin rate at which they can no longer be rubble pile bodies increases if there is some cohesive strength in the body. 

This paper probes the transition between rubble pile asteroids and their monolithic components, developing a theory for a physical process which may ``disaggregate'' rubble pile bodies efficiently into their monolithic components. To do this we jointly analyze the different effects which are known to dominate for small asteroids and explore the implications of their physics. Based on this analysis we develop a theory for the disaggregation of asteroids into their component parts that occurs over relatively short time spans (less than a few million years) and is a function of an asteroid's size, density and other physical characteristics. In it, we trace out a process by which small asteroids can be disaggregated from rubble pile bodies into their constituent aggregates, and provide a range of timescales over which this process can occur. The intent of the paper is to identify this process and its physical markers, enabling targeted observations to be made in order to check and constrain unknown parameters of our theory. We also discuss the application of this theory to asteroids of different types to probe differences that may exist between asteroids of different compositions. 

The paper is structured as follows. In Section 2 the basic physical theories that are combined together in this paper are discussed, summarizing their models and providing nominal values for key parameters. In Section 3 we combine these effects to develop specific predictions on the time scales and physical transition points that the physics suggest. In Section 4 the theory is applied to well characterized binary asteroids to develop strength and disaggregation time limits. In addition we make some notes on applying this theory to type sub-groups in the database. Based on our models, Section 5 provides a number of specific predictions and identifies a number of observations that should be pursued to better constrain and test the theory. 

\section{Background Physics and Models}

Physical descriptions and theory for the YORP effect, the strength of rubble pile bodies and conditions for a fissioned body to escape have been developed previously in the literature (briefly reviewed below). Here we summarize the main features of these models and define the controlling equations. 

\subsection{YORP Theory}

The YORP effect is by now a wholly accepted, if not fully understood, theory for how small asteroids can be subject to changing spin rates over time. 
The YORP effect for asteroids was first proposed by Rubincam \cite{rubincam_YORP} and later confirmed on a number of asteroids \cite{lowry_YORP, taylor_YORP, kaasalainen_YORP}. It describes the average effect of solar illumination on an asymmetric, rotating body. It can be shown that the averaged effect of YORP is to cause a uniformly rotating body to spin up or spin down over time. The effect has been probed down to small sizes, where at a size much less than a meter the physical manifestation of this effect has been predicted to become altered \cite{small_YORP}. At larger sizes, the YORP effect for a uniformly rotating asteroid can be described by a single coefficient which is a function of the solar inclination in the spinning body's equatorial reference frame \cite{ scheeres_YORP}. The action of YORP when the body is in a complex rotation state is more difficult to assess \cite{cicalo_YORP, breiter_tumbling_YORP}, but we will avoid this issue by assuming the body relaxes back to a uniform rotation state in a characteristic time which will be estimated. It is important to note that the YORP theory developed by Rubincam was based, in part, by earlier work by Paddack \cite{paddack1969} (the ``P'' in YORP), which was proposed as an effect that would cause small meteoroids to spin to disruption. In essence, the current study shows that Paddack's idea may be applied to smaller rubble pile asteroids. 

Following the notation and derivation of \cite{scheeres_YORP}, the YORP induced spin rate acceleration acting on a body is found to be
\beq
	\dot{\omega} & = & \frac{3\Phi}{4\pi A^2\sqrt{1-E^2}} \frac{{\cal C}}{\rho R^2} \label{eq:1}
\eeq
where ${\cal C} \sim 0.001 \rightarrow 0.01$ is a non-dimensional constant that is a function of the shape and obliquity of an asteroidal body \cite{scheeres_YORP}, the parameter $\Phi \sim 1\times 10^{17}$ kg-m/s$^2$ is the radiation constant, $A$ is the orbit semi-major axis and $E$ the eccentricity. 
Also of relevance is the YORP timescale, which is defined by the current spin rate $\omega$ divided by $\dot\omega$ 
\beq
\tau_Y & = & \frac{4\pi A^2\sqrt{1-E^2} }{3\Phi} \frac{\omega \rho R^2}{{\cal C}} \label{eq:TYORP}
\eeq
This is the time it takes an asteroid to double its current spin rate, or conversely to reduce its current spin rate to zero. 
We note that recent analysis has shown that the YORP effect can sometimes lead to stable equilibria in spin rate \cite{golubov2014three}, which could affect YORP timescales, and is an important topic for future research.

\paragraph{Relaxation Timescales}
One prediction of rubble pile fissions are that the component bodies of a fissioned rubble pile will immediately start to tumble, due to the conservation of the body spin vector across a breakup and the immediate mis-alignment of this spin vector from the principle moments of inertia of the body \cite{sanchez_MAPS}. While extensive study of the YORP effect on a tumbling body has not been made, it is believed that the YORP effect will be less significant for a tumbling body. Thus, the relaxation time of a body is important to consider. Taking the classical analysis found in Burns and Safronov  \cite{burns_safronov}, the relaxation timescale of a tumbling body back to uniform rotation about the maximum moment of inertia equals the ratio of the spin rate over the estimated rate of change in spin rate, 
\beq
	\tau_R & \sim & f \frac{\mu Q}{K_3^2 \rho R^2 \omega^3} \label{eq:tauR}
\eeq
where $f$ is a correction factor discussed below, $K_3^2 \sim 0.01 \rightarrow 0.1$ and is a function of the body's shape and the rigidity $\mu Q$ 
takes on values from $1.3\times10^7\rightarrow 2.7\times10^9$. The lower value has been recently constrained by observation of a body in the BYORP-tide equilibrium \cite{scheirich_FG3} while the upper number is based on constraining the tidal expansion rate of binaries \cite{taylor_margot_binaries}. More recent analyses by Pravec et al. \cite{pravec2014tumbling} indicate that the timescale should be reduced by a factor of 7-9, leading us to use a correction factor $f = 1/8$. Further, constraints on singly synchronous binary asteroids that are members of pairs also indicate that relaxation times may be short in general \cite{pravec2016binary}. 

To compare this with the YORP effect, divide the relaxation timescale by the YORP timescales to find
\beq
	\tau_R / \tau_Y & = & f \frac{3\Phi}{4\pi A^2\sqrt{1-E^2}} \frac{ \mu Q {\cal C}}{K_3^2} \frac{1}{\rho^2 R^4 \omega^4} 
\eeq
which will be evaluated later. 
Recent work has shown that the dissipation parameter $\mu Q$ may have a size dependence \cite{goldreich, jacobson_BYORP_equilibrium}, although we do not assess this potential effect. 

\subsection{Rubble Pile Strength}

Evidence for cohesion in rubble piles has been proposed recently based on theoretical predictions and observations. In \cite{scheeres_cohesion, sanchez_MAPS, sanchez_disruption} the physics and implications of cohesive forces on rubble pile bodies was explored. Predictions from the physics of the bodies, and from comparisons with lunar regolith material (which is the only space material to have been measured), indicate cohesion with a conservative range of values within 1 - 1000 Pa. Recent observations of intact and disintegrating asteroids also allows for cohesive strength limits to be placed on asteroids, with values being greater than 75 Pa for Asteroid 1950 DA \cite{rozitis2014cohesive, toshi_1950DA} and range between 50-100 Pa for P/2013 R3 \cite{hirabayashi_R3,2017arXiv170309668J}. Thus exploration of the implications of cohesion is a well motivated endeavor. 

For a rubble pile with a given level of cohesion the nominal failure rate of the body is a combination of overcoming gravitational attraction and cohesion. Holsapple \cite{holsapple_original, holsapple_plastic} has studied the conditions under which failure of a rubble pile will occur. S\'anchez and Scheeres \cite{sanchez_MAPS} have applied this theory to a general, elongate body to develop an approximate expression for failure spin rate,
\begin{eqnarray}
	\omega_F^2 & = & \omega_c^2 + \frac{2\sigma }{\rho\alpha^2} \frac{3-\sin\phi}{3+\sin\phi}  \label{eq:fission} \\
	\omega_c & = & \sqrt{\frac{4\pi{\cal G}\rho}{3}}\label{eq:omega_c}
\end{eqnarray}
where $\omega_F$ is called the failure spin rate and represents when the body will undergo plastic deformation, $\omega_c$ is the spin rate at which the asteroid begins to experience tensile forces at its surface, and is the rate where deformation definitely starts for a cohesionless body, $\sigma$ is the tensile, uniaxial strength that holds components of the body together and is measured in Pascals, $\rho$ is the density of the body, $\alpha$ is the longer radius of the body and $\phi$ is the angle of friction. The angle of friction for a granular asteroid is thought to lie between 30$^\circ$ and 45$^\circ$, giving the ratio $(3-\sin\phi)/(3+\sin\phi)$ a value ranging from $ 0.71 \rightarrow 0.62$. Taking an extreme value of $90^\circ$ just gives a value of 0.5, which we will adopt for conservatism in the following. 

From this result we can clearly see that for a given strength, a smaller body must spin more rapidly before it will fail. Conversely, we note that for a larger body, this term becomes small and the cohesionless spin limit will be larger in magnitude and cannot be ignored.
Due to the presence of strength in our model we will often only be considering bodies spinning considerably above the cohesionless spin limit, which will be relatively small compared to the cohesive component. Thus we use the approximation
\beq
	\omega_F & \sim & \sqrt{ \frac{\sigma}{\rho R^2}} \left[ 1 + \frac{1}{2} \frac{4\pi{\cal G}\rho}{3} \frac{\rho R^2}{\sigma} + \cdots \right] 
\eeq
also replacing the body's long axis with the mean radius $R$.  


\subsection{Fission and Conditions for Escape}

Fission and its role in the formation of asteroid pairs has been documented with clear evidence in Pravec \cite{pravec_fission} and its concordance with the underlying theory of fission by Scheeres \cite{scheeres_fission}. More generally, this process can yield cyclic systems, binary systems that are stable over long time periods, or can also lead to the persistent -- if not slow -- loss of mass from a rubble pile system \cite{jacobson_icarus}. A main way this occurs is through the formation of asteroid pairs, which requires the fissioned component of a body to satisfy a size constraint as a function of the initial body's morphology \cite{scheeres_F2BP, scheeres2017constraints}. 

In the simplest case for a body split into two components, the condition for these two components to be able to escape from each other is that the total energy of the system (including rotational energy of the individual bodies) be greater than zero, which was found to be consistent with the majority of detected asteroid pairs \cite{pravec_fission}. If the body splits at low or no cohesion (i.e., if the body is large enough so that this is not an important factor) the orbital energy will never be positive in general and escape can only result after a period of intense (but relatively brief) orbital evolution of the system where excess energy is transferred from the larger body's rotation state to the mutual orbit, eventually resulting in escape and the primary body spinning at a slower rate \cite{jacobson_icarus}. 

In fact, when bodies are subject to the fission conditions described above, there are many mechanisms that can occur, leading to the formation of (possibly long-lived) binary systems, or capture into a recurring cycle that conserves the system mass, or at most only allows small fractions of the body to escape after any given fission event. However, as a body that undergoes fission becomes inexorably smaller, the cohesion that exists within a rubble pile body causes the spin rate for fission to increase, and hence the total orbital energy of the fissioned system will increase. One outcome is that for smaller bodies, the mass fraction of the system that will have a positive total energy will increase, and thus this could allow for bodies with a larger mass fraction to form asteroid pairs. For even smaller bodies, however, there will come a point where the body spins fast enough at failure so that the orbital energy of the fissioned system is positive, meaning that the system components will immediately escape and will not have a period of orbital interaction. We are ultimately interested in this transition point, as once a body is small enough for this to hold its evolution will no longer be subject to the complex evolution described above \cite{jacobson_icarus}, but instead the fissioned components will abruptly escape from each other. This point also defines where we expect binary asteroids to no longer form, since the characteristic interplay between the components implicated in the formation of binaries can no longer occur \cite{binary_AIV}. 

To derive the spin limit for a positive orbital energy we use a simple model consisting of two spheres resting on each other, with radii $R_1$ and $R_2$. Assuming a common density, the total mass of the system will be $\frac{4\pi}{3} \rho \left(R_1^3 + R_2^3\right)$, the distance between the mass centers will be $R_1+R_2$, and once failure occurs the relative velocity between the components will be $\omega(R_1+R_2)$. Putting this into the orbital energy equation and making it greater than zero yields
\beq
	\frac{1}{2} \omega^2 \left(R_1 + R_2\right)^2 - \frac{4\pi {\cal G} \rho}{3}\frac{ \left(R_1^3 + R_2^3\right)}{R_1 + R_2} & > & 0
\eeq
where ${\cal G}$ is the gravitational constant. Solving for a condition on the spin rate we find
\beq
	\omega^2 & > & \frac{8\pi {\cal G} \rho}{3}\frac{ \left(R_1^3 + R_2^3\right)}{\left(R_1 + R_2\right)^3}
\eeq
For this system, the mean radius of the parent body is $R^3 = R_1^3 + R_2^3$, and thus we can introduce the scaled radii $r_1 = R_1 / R$ and $r_2 = R_2 / R$, with the constraint $r_1^3 + r_2^3 = 1$. The right hand side then has the functional form $\frac{1}{\left( r_1 + r_2\right)^3}$.  It can be easily shown that this factor takes on a minimum value of 1/4 and a maximum of 1. 

To be consistent with our failure theory, which does not account for the lower spin rates needed for a bimodal shape model, and for conservatism, we take the upper limit for escape, thus defining the criterion for when a body will start to disaggregate as
\beq
	\omega_E & = & \sqrt{\frac{8\pi {\cal G} \rho}{3}}
\eeq
or $\sqrt{2} \omega_c$. The lower limit for positive energy occurs when the bodies are equal in size, and yields a spin rate for abrupt escape of $\omega_c / \sqrt{2}$, but would also require us to change the failure criterion. 

\subsection{Mass Conservation}

To proceed with our analysis requires that we develop a rule for how a rubble pile body will split into its components. This is a question of interest in general and has been studied using a variety of techniques \cite{holsapple_YORP, sanchez_icarus}. Our main concern is that the process conserve mass and provide a simple and recursive result that can be scaled across arbitrary sizes. This is a clear simplification, and indeed it may be expected that the manner in which a rubble pile component will split may change as a function of size. Consideration of such effects is left for future studies, however. A key component of our model is that the body splits into components of finite size relative to each other. This is to be expected based on the mechanics of fission \cite{PSS_fission} and conforms with recent investigations of the fission of cohesive rubble pile bodies \cite{sanchez_disruption}. 

With the main goal of mass conservation across such an abrupt fission event, we develop a rule for relating the size of the new asteroids with its parent. If the body breaks into $N$ distinct components, then on average $N R_{i+1}^3 = R_i^3$, where $R_i$ is the mean radius of the body that splits. This also provides a means to link the size of a body through a number of different fissions. Applying this rule, we can define the size reduction at each step, and if we assume that there are $i$ fission events starting from an initial size $R_0$ we have
\beq
	R_{i} & = & \frac{1}{N^{1/3}} R_{i-1} \\
	R_{i} & = & \frac{1}{N^{\frac{i}{3}}} R_0
\eeq
This simple rule, with its free parameter $N$, will be used throughout the analysis. 

\section{Combination of Effects}

Given the different physical elements, as described above, we now combine them into specific relationships to predict the size at which a body will begin an abrupt escape phase upon failure, and estimates of the time between such fissions once it has entered this phase. 

\subsection{Size for Abrupt Escapes}

First compare the failure spin rate, $\omega_F$, with the condition for escape, $\omega_E$. Note that for defining the start of this process we use the full failure spin rate formula. In order for the process of abrupt escape to start requires that $\omega_F > \omega_E$. Using the squares of these quantities, this yields
\beq
	\frac{4\pi{\cal G} \rho}{3} + \frac{\sigma}{\rho R^2} & > & \frac{8\pi {\cal G} \rho}{3}
\eeq
which leads to the condition on $R < R_{0}$, where
\beq
	R_0 & = & \frac{1}{ \rho}\sqrt{ \frac{3\sigma}{4\pi {\cal G}}  } \label{eq:R0}
\eeq
The size $R_0$ then defines when we would expect a specific asteroid to enter its disaggregation phase and will also serve as a lower limit on binary asteroid sizes. 

\subsection{Time Between Fission Events}

To estimate the time between fission events we must bring together several of the above effects. 
First, in Eqn.\ \ref{eq:fission} we note that the cohesive term is in general larger than the cohesionless term when the body size $R < R_0$. Thus, as the body continues to disaggregate we can gauge how fast this term grows relative to the cohesionless term as the bodies become smaller. If these terms are equal at $R_0$, then the cohesive term increases by a factor $N^{2/3}$ for each new generation. Table \ref{tab:1a} shows these values as a function of $N$. Noting the significant increase of the cohesive term even over a few generations, we justify treating the cohesionless term as small in the following, allowing for the simplification
\beq
	\omega_F & \sim & \sqrt{ \frac{\sigma}{\rho R^2}   } + \frac{2\pi{\cal G}\rho \sqrt{\rho} R}{3 \sqrt{\sigma}} + \ldots 
\eeq

\begin{table}[htb]
\centering
\caption{Increase factor per generation as a function of the number of components.}
\label{tab:1a}
\begin{tabular}{| c | c | c | c | c | c | c |}
\hline
$N$ & 2 & 3 & 4 & 5 & 6 \\
\hline
$N^{2/3}$ & 1.59 & 2.08 & 2.52 & 2.92 & 3.30 \\
\hline
\end{tabular}
\end{table}

Then, in general, the time it takes a body to spin from a zero spin rate up to the failure spin rate $\omega_F$ equals the failure spin rate divided by the YORP acceleration from Eqn.\ \ref{eq:1}, yielding in general
\beq
	\tau_F & = & \frac{4\pi A^2\sqrt{1-E^2} }{3\Phi} \frac{R \sqrt{\rho \sigma}}{{\cal C}} 
\eeq
where we completely neglect the cohesionless term for the moment. 

Using these results we can compute the time it takes to go from one fission state to the next. We adopt two approaches, one is to compute the shortest time possible, essentially assuming that the components continue to spin up in the same direction until they hit the next fission spin rate. The other is to assume the bodies spin up in the opposite direction, pass through zero, and then spin up to the next fission rate. These two computations will bracket the timespan required until the subsequent fission. It is important to account for possibly significant changes in the YORP coefficient following fission or reconfiguration of the asteroid. The sensitivity of a body's YORP coefficient has been well documented \cite{scheeres_itokawa_YORP} and studied in some detail \cite{statler}. Some theories show a continuous coupling between changes in spin rate and the YORP coefficient \cite{cotto_statler_YORP}, however these models do not account for plastic deformation of the body and do not include cohesive forces, which are expected to be important for smaller rubble pile bodies, and thus are not accounted for here. 

\paragraph{Consistent Spin-Up Direction} To calculate the time between fissions in this case, directly difference the fission spin rate $\omega_{F, i+1}$ with the prior rate $\omega_{F,i}$ to compute
\beq
	\Delta \omega_{F,i+1} & = & \sqrt{\frac{\sigma}{\rho}} \left[ \frac{1}{R_{i+1}} - \frac{1}{R_i} \right] \\
	& = & \sqrt{\frac{\sigma}{\rho}} \left[ 1  -  \frac{1}{N^{1/3}} \right] \frac{1}{R_{i+1}}
\eeq
The time to cover this change in spin rate is found by dividing by $\dot{\omega}$ in Eqn.\ \ref{eq:1} evaluated for a body of radius $R_{i+1}$, as above, to find
\beq
	\Delta t_{i+1} & = & \frac{4\pi A^2\sqrt{1-E^2} }{3\Phi} \frac{\sqrt{\rho \sigma}}{{\cal C}} R_{i+1} \left[ 1  -  \frac{1}{N^{1/3}} \right] 
\eeq
Substituting in for our general model of $R_{i+1}$ as a function of the initial radius $R_0$ then yields
\beq
	\Delta t_{i+1} & = & \frac{4\pi A^2\sqrt{1-E^2} }{3\Phi} \frac{\sqrt{\rho \sigma}}{{\cal C}} R_{0} \left[ 1  -  \frac{1}{N^{1/3}} \right] \left(\frac{1}{N^{1/3}}\right)^{i+1} \label{eq:deltat}
\eeq

To compute the total time $T_{i+1}$ it takes to go from the initial fission at size $R_0$ to the current $i+1$ generation, we compute the summation of the times, $T_{i+1} = \sum_{j =1}^{i+1} \Delta t_{j}$ which can be conveniently expressed as
\beq
	T_{i+1} & = & \frac{4\pi A^2\sqrt{1-E^2} }{3\Phi} \frac{\sqrt{\rho \sigma}}{{\cal C}} R_{0} \left[ 1  -  \frac{1}{N^{1/3}} \right]  \sum_{j=1}^{i+1} \left(\frac{1}{N^{1/3}}\right)^{j}
\eeq
However, we recognize the summation as a simple power series, with the partial sum $\sum_{j=1}^{i+1} \epsilon^j = (1-\epsilon^{i+2}) / (1 - \epsilon)$, which in our case yields
\beq
	T_{i+1} & = & \frac{4\pi A^2\sqrt{1-E^2} }{3\Phi} \frac{\sqrt{\rho \sigma}}{{\cal C}} R_{0} \left[ 1 -  \left(\frac{1}{N^{1/3}}\right)^{i+2} \right]
\eeq
If we take the limit as $i\rightarrow\infty$ we find the lifetime to be finite and equal to 
\beq
	T_{\infty} & = & \frac{4\pi A^2\sqrt{1-E^2} }{3\Phi} \frac{\sqrt{\rho \sigma}}{{\cal C}} R_{0}
\eeq
Finally, replace $R_0$ with the expression from Eqn. \ref{eq:R0} to find
\beq
	T_{\infty} & = & \sqrt{\frac{4\pi}{3}} \frac{A^2\sqrt{1-E^2} }{\Phi} \frac{\sigma}{{\cal C} \sqrt{{\cal G}\rho}} \label{eq:tinf}
\eeq

There are a few interesting elements to point out. First, for the consistent spin up direction we find that the lifetime is independent of the number of components the body fissions into. Second, we see that the time is proportional to the strength of the body, thus a stronger rubble pile will take proportionately longer to spin to full disaggregation (this in fact is driven by the original body $R_0$ being larger as well). Next, the time is inversely proportional to the square root of $\rho$. This is a bit counterintuitive, and implies that rubble piles with lower densities will take longer to disaggregate, and those with larger densities take a shorter time. This arises due to the competition between the YORP timescale, where a more dense body will take longer to spin up, versus the failure spin rate, which decreases with increasing density due to the inertial forces which pull the body apart, and to the initial body radius $R_0$, which is smaller for a larger density. Third is that the lifetime equals the YORP timescale for a body of radius $R_0$ divided by $\sqrt{2}$. This can be seen by substituting the appropriate values into Eqn.\ \ref{eq:TYORP}. 

Thus, for a consistent spin up we find that the lifetime once abrupt escape begins is $T_{\infty}$, given in Eqn.\ \ref{eq:tinf}. If we wish to add to this the overall YORP timescale of the body, accounting for the time it would take to initially spin a body of size $R_0$ up to this rate, the total disaggregation lifetime then equals $(1+\sqrt{2}) T_{\infty}$. 

\paragraph{Opposing Spin Direction}

Next, let us apply the same analysis, but at each step allow for the body to first be spun down to a zero spin rate and then spun up to disruption in the opposite direction. 
This simple model sidesteps known issues with how an asteroid's rotational dynamics evolves when it gets to a slow spin rate. Previous investigations have detected a statistically larger number of slowly rotating asteroids \cite{pravec2008spin,rossi_rotation}, which could be a signature of more complex rotational evolution and a slow-down in the rate at which a body's spin rate varies \cite{vokrouhlicky_tumbling,cicalo_YORP,breiter_tumbling_YORP}. We do not pursue this topic here, however, and instead use a simple model which could be adjusted by a scale factor, once a consistent theory for the evolution of a slowly-rotating and tumbling asteroid is developed. 

The analysis is much the same as before, however at each step $i+1$ we must add twice the YORP timescale for a body of radius $R_{i+1}$ at a spin rate of $\omega_{F,i}$. Thus, to every term $\Delta t_{i+1}$ in Eqn.\ \ref{eq:deltat} we must add
\beq
	2 T_{\infty} \left( \frac{1}{N^{1/3}} \right)^{i+2} 
\eeq
where the definition of $T_{\infty}$ in Eqn.\ \ref{eq:tinf} has been used to simplify the expressions. Summing this expression from $j = 1, i$ yields a convergent power series again, and equals 
\beq
	2 T_{\infty} \frac{1 - \frac{1}{N^{(i+1)/3}}}{N^{1/3} - 1} 
\eeq
Letting the limit $i \rightarrow \infty$ and adding to the original lifetime $T_{\infty}$ we get the disaggregation lifetime once the body enters the abrupt escape phase as
\beq
	T_{\infty}' & = & T_{\infty} \frac{N^{1/3} + 1}{N^{1/3} - 1}
\eeq
which is just a correction to the consistent spin up direction lifetime. If we wish to account for the initial spin up of $R_0$ to its failure rate then we just add $\sqrt{2} T_{\infty}$ again. A few additional comments can be made about this term. First, we see that as the body breaks into more components, or as $N \gg 2$, that this time approaches the ``consistent direction'' time. Second, we can evaluate the lifetime extension as a function of how many components the body fissions into, with some specific computations given in Table \ref{tab:2}. We note that the lifetime extension is always less than an order of magnitude, but that it decreases somewhat slowly as $N$ increases. 

\begin{table}[htb]
\centering
\caption{Increase factor for spinning in the opposite direction as a function of the number of components a body breaks into.}
\label{tab:2}
\begin{tabular}{| c | c | c | c | c | c | c | c |}
\hline
$N$ & 2 & 3 & 4 & 5 & 10 & 100 \\
\hline
& & & & & &  \\
$\frac{N^{1/3}+1}{N^{1/3}-1}$ & 8.69 & 5.52 & 4.40 & 3.82 & 2.73 &  1.55 \\
& & & & & &  \\
\hline
\end{tabular}
\end{table}

\subsection{Relaxation Time Effects}

We can also account for relaxation time, where we may assume that the YORP effect will not become strong again until the body has settled into near principal axis rotation. Although this phenomenon is not fully understood, both for the relaxation time and the response of an asteroid to YORP when in a complex rotation state, we add this discuss for conservatism. 

Let us compute the relaxation time when the body splits from size $R_i$ to $R_{i+1}$. The spin rate will then be at $\omega_{F,i}$ and the size at $R_{i+1}$. Inserting these values in Eqn.\ \ref{eq:tauR} gives the relaxation time following this split as
\beq
	\tau_{R, i+1} & = & f \frac{\mu Q}{K_3^2} \sqrt{\frac{3}{4\pi{\cal G}}} \frac{N}{\sigma \sqrt{\rho}} \frac{1}{N^{(i+1)/3}}
\eeq

Summing over all the splits again, and letting the summation go to $\infty$, we find the total relaxation time is  
\beq
	T_R & = & f \frac{\mu Q}{K_3^2} \sqrt{\frac{3}{4\pi{\cal G}}} \frac{1}{\sigma \sqrt{\rho}} \frac{N}{1 - 1 / N^{1/3}}
\eeq
Here we see different dependencies relative to the time $T_{\infty}$. First, as the body becomes stronger or more dense we see that the time shortens. This seems counterintuitive, but is driven by the fact that a stronger body must spin faster in order to fission, which also decreases its relaxation time. Second, we note that when the bodies disaggregate into a large number of components, $N$, that the lifetime increases significantly. This is due to a smaller body having a longer relaxation time in general. 

For context, we divide by $T_{\infty}$ to gauge the significance of this term relative to the YORP effect. Doing so yields 
\beq
	T_R / T_{\infty} & = & f \frac{3}{4\pi} \frac{\mu Q}{K_3^2} \frac{\Phi}{A^2\sqrt{1-E^2}} \frac{{\cal C}}{\sigma^2} \frac{N}{1 - 1 / N^{1/3}} \label{eq:TR_Tinf}
\eeq
The main points of interest here is that the relative time is independent of the density and it varies inversely with strength squared. 

\subsection{Correction for Cohesionless Spin}

Finally, we can evaluate the error introduced in our analysis by neglecting the contribution of $\omega_c$ to the overall spin rate of the body. By performing a higher order expansion one can show that the true failure spin rate must be accelerated above the regular formula when accounting for the cohesionless term. Specifically, the first order correction for a body of radius $R_i$ becomes
\beq
	\Delta \omega_{F,i} & = & \frac{2\pi{\cal G}}{3} \sqrt{\frac{\rho^3}{\sigma}} R_i
\eeq
Thus we see that this contribution reduces linearly with the radius of the $i$th body, and becomes less and less important. The additional time that must be added per fission is found by dividing this rate by the YORP acceleration, evaluated at a body size $R_i$. Carrying out this division we find
\beq
	\Delta t_{F,i} & = & \frac{1}{2} \left( \frac{4\pi}{3}\right)^2 \frac{A^2\sqrt{1-E^2}}{\Phi} \frac{\rho^{5/2}}{\sqrt{\sigma}{\cal C}} R_i^3
\eeq
Summing this over all contributions gives us the additional time that should be accounted for. This can be conveniently expressed in terms of the disaggregation time $T_{\infty}$ as
\beq
	\Delta T_{F} & = & \frac{1}{2} T_{\infty} \frac{1}{1 - 1/N} 
\eeq
We see that the time is at most doubled when $N=2$, and asymptotically approaches a 50\% increased time for $N\gg 1$. 

\subsection{Limits on Disaggregation Lifetime}

Now the different computations and corrections to the disaggregation times derived in this Section are combined into limits on the disaggregation time, denoted as $T_D$. The lower limit is given by the direct spin up case and the upper limit by the reversing spin up case. To both we add the correction $\Delta T_F$ derived above. We do not add the relaxation time addition, but will evaluate the range of values for this term in the following. We also do not add in the time estimate for the initial body of radius $R_0$ to be spun up to its failure rate, which initiates the disaggregation process. 

Then the theory predicts that the disaggregation time for a body of radius $R_0$ is bounded between
\beq
	\frac{3}{2} T_\infty \frac{\left( 1 - \frac{2}{3N}\right)}{\left(1 - \frac{1}{N}\right)} & \le T_D \le & 
	\frac{3}{2} T_\infty \frac{ 1 + \left( 1 / N^{1/3} - 2 / N - 2 / N^{4/3}\right) / 3}{\left[1 - 1/N^{1/3}\right]\left[1 - 1/N\right]}
\eeq

These results end up having a somewhat complex relationship with the number of components the body breaks up into, $N$. Thus we give the limits for two special cases, $N = 2$ and $N \gg 2$. 
If the body consistently splits into two components we find
\beq
	2 T_\infty & \le T_D \le & 
	9.69\ldots  T_\infty 
\eeq
If instead the body splits into $N \gg 2$ components we have the approximate results 
\beq
	\frac{3}{2} T_\infty \left[ 1 + \frac{1}{3N} + \ldots\right] & \le T_D \le & 
	\frac{3}{2}  T_\infty \left[ 1 + \frac{4}{3 N^{1/3}} + \ldots \right]
\eeq
For this result, note that the $1/N$ term at the lower limit is negligible with respect to the $1/N^{1/3}$ term on the upper limit and can be ignored at the given order of accuracy. The interesting item to note again is that as $N$ grows large these two limits converge to $3/2 T_\infty$, which is shorter than the limiting time for disruption into two equal components. 

\section{Bounds on the Strength and Disaggregation Lifetime of Rubble Pile Asteroids}

The theory is now applied to known binary asteroids and the binary asteroid population in general in order to develop upper and lower bounds on the strength of these rubble piles, and their corresponding disaggregation lifetimes. Considering binary asteroids provides us with an estimate of the density of the body. This, in conjunction with the size of the primary is then used to develop an upper limit on the body's strength. Using this value, in conjunction with other parameters discussed shortly, we can also find upper bounds on the disaggregation lifetime of the body. We consider the smallest binary asteroids for which we have a spectral type classification in order to develop the most restrictive bounds on asteroids in these populations. Independent of asteroid type, we also develop a lower bound by choosing a minimum binary size that is less than all binaries discovered to date, and consider the corresponding strength limits for different density assumptions. In this way we bracket the strength of rubble pile asteroids and their disaggregation lifetime. 

For the binary asteroid populations we use the current best estimates for the asteroid binary population physical data from the Ondrejov Observatory database\footnote{http://www.asu.cas.cz/$\sim$asteroid/binastdata.htm, update 2015 09 18} \cite{pravec2016binary} and for the asteroid spectral types from the Near-Earth Asteroids Data Base\footnote{http://earn.dlr.de/nea/table1\_new.html, update Mon Apr 03 17:42:09 2017} \cite{binzel2002physical}. Thus, our analysis can be easily updated in the future as more observations are accumulated and binary asteroids discovered. We also note that the binary size data is given in terms of diameter, thus our specific implementations switch to diameter instead of radius, or $R = D/2$. 

For the disaggregation time calculations there are important physical parameters beyond the bulk density which must be assumed. These are the YORP coefficient ${\cal C}$ and the dissipation parameters $\mu Q$ and $K_3^2$. Also of relevance is the number of components a body will split into, $N$, although this is just a model parameter. 

For the YORP coefficient we rely on the observations made in \cite{scheeres_YORP}, which performed a limited survey of the YORP coefficients of a number of different asteroid shapes. From that study it was found that most bodies have a non-dimensional YORP coefficient in the range $0.001 \le {\cal C} \le 0.01$, which provides for an order of magnitude variation in this parameter. A useful item for future study would be a better characterization of the YORP coefficient across asteroid shapes. 

We must also make assumptions for the dissipation parameter values, which are quite uncertain in general. Here for the shape parameter we adopt $K_3^2 \sim 0.1$ which is the value from \cite{harris_relax} for oblate bodies. For the parameter $\mu Q$ we consider a range of values going from the inferred value of $2.7\times 10^9$ Pa based on population studies \cite{taylor_margot_binaries}, to a lower value of $1.3\times 10^7$ Pa based on direct measurement of parameters from the binary asteroid 1998 FG3 \cite{scheirich_FG3}, which is presumed to be in a BYORP-Tide equilibrium \cite{jacobson_BYORP_equilibrium}.

\subsection{Evaluation Procedure}

For a given binary asteroid we use the estimated density of the primary body to supply the bulk density needed. The size of the primary, $D$, also provides an upper limit on the disaggregation size, $D_0$, defined above. With these two determinations we then find an upper bound on the strength of the asteroid to be 
\beq
	\sigma & \le & \frac{\pi {\cal G}}{3} \left( \rho D\right)^2 
\eeq

Given a strength value we can consider the range of values for the predicted disaggregation time parameter $T_\infty$. Expressing the binary orbit in terms of AU, the value of the parameter is
\beq
	T_\infty & = & 5.6 \times 10^{10} A^2 \sqrt{1-E^2} \frac{\sigma}{{\cal C}\sqrt{\rho}} \mbox { s } \\
	& = & 1.8\times 10^{3} A^2 \sqrt{1-E^2} \frac{\sigma}{{\cal C}\sqrt{\rho}} \mbox { yr } 
\eeq

Finally, we compare the additional time for relaxation with the disaggregation time $T_\infty$, from Eqn.\ \ref{eq:TR_Tinf}. Taking $N = 2$ for definiteness, we find
\beq
	T_R/T_\infty & = & 1.3 \times 10^{-5} \frac{1}{A^2 \sqrt{1-E^2} } \frac{ \mu Q {\cal C}} {\sigma^2} 
\eeq
We can consider the different ranges of these values separately for our different binary asteroid examples. 

For our lower bounds, we note that no binary asteroid has been found of size lower than 100 m. This despite a relatively rich set of asteroid populations at sizes smaller than this limit \cite{rotation_database}. The other controlling parameter is the bulk density, and here we can consider typical densities associated with different asteroid complexes.

\subsection{Minimum-size Binary Asteroids for different Types}

In Table \ref{tab:1} we list the binary asteroids considered. The relevant information for each system is given, and the interested reader is directed to the Ondrejov Observatory database for additional details. We do not take the full range of uncertainties from that database, although this could be done. However, as these lifetimes and strengths are upper bounds for the different asteroid types, we do not find it useful to add yet another layer of ambiguity to the calculations. From the data the overall minimum binary asteroid size is 120 m for asteroid 2003 SS84. This asteroid has no type designation, thus serving as an ambiguous lower bound on the population. For added conservatism we can take a lower size of 100 m. The change in strength by using the lower size is on order of 44\%. 

For the $T_\infty$ and $T_R/T_\infty$ computations we use the larger value of the YORP coefficient, ${\cal C}$, noting that it can decrease by an order of magnitude. For the parameter $\mu Q$ we take the larger value, noting that it can decrease by over 2 orders of magnitude. Thus, $T_\infty$ can increase by an order of magnitude and $T_R/T_\infty$ can decrease from the values in the table by 3 orders of magnitude.

\begin{landscape}

\begin{table}[htb]
\caption{Minimum size binary asteroids for upper and lower bounds on strength and disaggregation time. We list all type designations that are given in the Near-Earth Asteroids Database, yet also loosely group them into different complexes.}
\label{tab:1}
\begin{tabular}{| r l || c | c | c | c || c | c | c |}
\hline
\multicolumn{2}{|c||}{Asteroid} & Type & Complex & $\rho$ & $D$ & $\sigma$ & $T_\infty$ & $T_R / T_{\infty}$ \\
  &  & & & (kg/m$^3$) & (m) & (Pa) & (yr) &  \\
\hline
\hline
\multicolumn{9}{|c|}{\bf Upper Bound} \\
\hline
481532&	2007 LE&	C&	C/B	&2000	&500&	6.987E+01&	2.922E+05&	6.914E-02 \\
175706&	1996 FG3&	Xc;B;Ch&	C/B	&1300	&1640&	3.176E+02&	1.647E+06&	3.346E-03\\
3671&	Dionysus&	Cb;X&	C/B	&2000	&1430	&5.715E+02&	2.390E+06&	1.033E-03\\
\hline
	&2007 DT103&	Sq/Q	&S/Q	&2000	&300	&2.515E+01&	1.052E+05&	5.335E-01\\
85938&	1999 DJ4&	Sq&	S/Q	&2000	&350	&3.424E+01&	1.432E+05&	2.879E-01\\
399774&	2005 NB7&	S/Sr&	S/Q	&2000	&500&	6.987E+01&	2.922E+05&	6.914E-02\\
185851&	2000 DP107&	Sq&	S/Q	&1300	&860	&8.733E+01&	4.529E+05&	4.425E-02\\
136617&	1994 CC&	Sq;Sa&	S/Q	&2000	&620	&1.074E+02&	4.492E+05&	2.924E-02\\
31345&	1998 PG&	Sq;Q	&S/Q	&2000	&820	&1.879E+02&	7.858E+05&	9.557E-03\\
162483&	2000 PJ5&	Q&	S/Q	&2000	&820	&1.879E+02&	7.858E+05&	9.557E-03\\
385186&	1994 AW1&	Sa&	S/Q	&2000	&900	&2.264E+02&	9.466E+05&	6.586E-03\\
153958&	2002 AM31&	Q&	S/Q	&4300	&450	&2.616E+02&	7.460E+05&	4.932E-03\\
1862&	Apollo&	Q&	S/Q	&2000	&1550	&6.714E+02&	2.808E+06&	7.486E-04\\
\hline
363599&	2004 FG11&	V&	V	&2000	&150	&6.288E+00&	2.629E+04&	8.535E+00\\
5381&	Sekhmet&	V&	V	&1800	&1000	&2.264E+02&	9.978E+05&	6.586E-03\\
164121&	2003 YT1&	V,R,Sr&	V	&2000	&1000	&2.795E+02&	1.169E+06&	4.321E-03\\
\hline
65803&	Didymos&	Xk&	X	&2000	&750	&1.572E+02&	6.573E+05&	1.366E-02\\
\hline
\hline
\multicolumn{9}{|c|}{\bf Lower Bound} \\
\hline
 &  & & C/B & 1300 & 100 & 1.181E+00	& 6.124E+03	& 2.421E+02 \\
\hline
 &  & & S/Q & 2000 & 100 & 2.795E+00	& 1.169E+04	& 4.321E+01 \\
\hline
\end{tabular}

\end{table}

\end{landscape}

First consider the upper and lower bounds on rubble pile strength in Table \ref{tab:1}. For the C/B complex the minimum lowest upper bound is 70 Pa for the C type asteroid 2007 LE, while the well determined binary 1996 FG3 gives a much larger bound, over 300 Pa. Across the S/Q complex the minimum bound is 25 Pa for 2007 DT103. The V complex has a lower upper bound of only 6 Pa for 2004 FG11. On the other hand, taking the minimum size binary across all types as 100 m, we get a lower strength of 1 Pa for a density of 1300 kg/m$^3$ and about 3 Pa for 2000 kg/m$^3$. The upper bounds are consistent with other estimates of cohesive strength that have been developed, using similar theory. Most recently, Jewitt et al. \cite{2017arXiv170309668J} developed a strength estimate of 50-100 Pa for the active asteroid P/2013 R3 (still of uncertain type)  assuming a density of 1000 kg/m$^3$ (we note that the strength would increase with a larger density), while Hirabayashi and Scheeres \cite{toshi_1950DA} developed a lower limit on the strength of 1950 DA of 70 Pa, which is listed as a EM type in the EARN database. These comparisons show cohesive strengths that lie between a few and a few hundred Pa. According to the theory of rubble pile cohesion \cite{sanchez_MAPS}, this bulk strength is strongly dependent on minimum grain size, porosity and mineralogy. Thus, there is likely variation across asteroid types and even within complexes due to different morphological structures. 

The minimum values of $T_\infty$ across each complex are on the order of a few hundred thousand years or less. This leads to disaggregation times that are a factor of 1.5 or up to a factor of 10 longer. We note that $T_\infty$ scales with strength, thus the weakest bodies have very short disaggregation times. Competing with this, however, is the relaxation time which becomes larger for weaker bodies. Still, as weaker bodies have shorter lifetimes, the additional time for relaxation is still relatively short. Consider asteroid 2004 FG11, the lowest upper bound on strength with a $T_\infty$ of 26,000 years and the largest ration of relaxation time to $T_\infty$ of 8.5. Adding the relaxation time to $T_\infty$ gives a total time of under 250,000 years.

\section{Discussions and Implications}

Now a number of implications of this theory are explored, which provide a future means to test it and to motivate new observation interpretations. 

\paragraph{Meteoroid Implications}

We note that for any level of cohesion, asteroids are expected to enter a disaggregation phase once their diameter is reduced below $D_0$. While the time for a given rubble pile to shed components or lose mass to get below this size can be quite long, once this occurs we note that disaggregation times are generally short, lasting less than a few Myr in general. As disaggregation times decrease as asteroids move into the inner solar system, there should be more and more rubble pile asteroids being disaggregated into their fundamental components. Thus, this would predict that meteoroids that encounter the Earth may preferentially be single, monolithic components of a parent rubble pile that no longer exists. Further, this would indicate that the meteoroid population would be representative of the size distribution of boulders and grains within rubble pile bodies. 

Stronger rubble piles could have lifetimes long enough to ensure that they do not become completely disaggregated before impacting the sun or one of the inner planets. However, ``stronger'' here must be taken with a grain of salt, as even a body with a strength of a few hundred Pa is still extremely weak and would become disaggregated at high altitude if entering the Earth's atmosphere. This general trend could provide a discriminant between meteorites that 
fall as solid objects and which could have come from a weaker rubble pile that had previously disaggregated, and those that  
come from a stronger rubble pile that has not yet disaggregated and may fail at high altitudes. If specific trends regarding meteorite type and 
where they are seen to fail is available, this could provide insight on the cohesive strength of rubble piles of these respective types. 

\paragraph{General Size Distribution Evolution}

Once the disaggregation phase starts, a given body of size $D$ will ultimately be transformed into a size distribution defined by its rubble pile structure. This is analogous to the transformation of two impacting asteroids into a size distribution of debris resulting from their destruction. Incorporation of this approach into standard population models can be made and should be explored for the smallest components of the asteroid population. 

This effect could also contribute to the zodiacal dust, as it provides a method for bodies to disaggregate into their fundamental constituents. We should note that at the end game, when the final blocks are rapidly rotating yet may still have some regolith covering, to release $\sim 100$ micron grains from the surface of a boulder requires spin rates that are extremely high. At this small scale limit we can question whether the simple rules developed here will still apply. 

\paragraph{Small Asteroid Limit}

We also note that as a body becomes small it has been shown that the YORP effect goes to a $1 / R$ dependence \cite{small_YORP}. This means that the time between fissions no longer decreases once a body is sufficiently small, but should approach a constant value. Thus in reality the end game at small sizes should flatten out. This occurs at such a small size, however, that we essentially ignore this effect for our model. If it does occur, then the limiting time to disaggregation becomes a constant (replacing $1/R^2$ by $1/R$ in Eqn.\ \ref{eq:1} and carrying out the summation). Thus, depending on how many components the rubble pile breaks up into once disruption rates are reached at a given size, the disaggregation time will be proportional to the number of components in the rubble pile. 

\paragraph{Future Work}

We finally note that the range of possible values for disaggregation time and rubble pile strength are relatively large and should be reduced for the theory to have a sharper application. Major uncertainties currently exist in the modeling of the YORP effect, the fission characteristics of cohesive asteroids and in the relaxation time for complex rotators. 
On the observational side, it would be of interest to probe the binary cutoff size across asteroid populations of different mineralogies. 

\section{Conclusions}

It is shown that small levels of cohesion within a rubble pile asteroid will, when combined with the YORP effect, cause rubble pile asteroids to disaggregate into their fundamental constituents within relatively short time spans, on the order of a million years or potentially much less. The size at which this disaggregation phase starts and the time for disaggregation are functions of the body density, strength, and other geophysical parameters. 
The theory is applied to well characterized binary asteroids to find upper and lower limits of strength as a function of density and spectral type. Lower bounds are at a few Pascals while upper bounds are generally less than a few hundred Pascals, in some cases less than a few tens of Pascals.

\section*{Acknowledgements}

The author is grateful to Dr. Alessandro Rossi for his support at IFAC in Florence while writing this paper and for his help with interpreting the asteroid database. 
The author acknowledges support from NASA grant NNX14AL16G from the Near Earth Objects Observation programs and from NASA's SSERVI program. 
The useful comments and criticisms of the referees Petr Pravec and Alan W. Harris are acknowledged. 

\bibliographystyle{plain}
\bibliography{../../../bibliographies/biblio_conferences,../../../bibliographies/biblio_article,../../../bibliographies/biblio_books,../../../bibliographies/biblio_misc}

\begin{thebibliography}{10}

\bibitem{binzel2002physical}
Richard~P Binzel, Dmitrij~F Lupishko, Mario Di~Martino, Richard~J Whiteley, and
  Gerhard~J Hahn.
\newblock Physical properties of near-earth objects.
\newblock {\em Asteroids III}, 255:271, 2002.

\bibitem{bottke_AIV}
William~F Bottke, Miroslav Bro{\v{z}}, David~P O?Brien, Adriano~Campo Bagatin,
  Alessandro Morbidelli, and Simone Marchi.
\newblock The collisional evolution of the main asteroid belt.
\newblock {\em Asteroids IV}, pages 701--724, 2015.

\bibitem{breiter_tumbling_YORP}
S~Breiter, A~Ro{\.z}ek, and D~Vokrouhlick{\`y}.
\newblock Yarkovsky--o'keefe--radzievskii--paddack effect on tumbling objects.
\newblock {\em Monthly Notices of the Royal Astronomical Society},
  417(4):2478--2499, 2011.

\bibitem{small_YORP}
S~Breiter, D~Vokrouhlick{\`y}, and D~Nesvorn{\`y}.
\newblock Analytical yorp torques model with an improved temperature
  distribution function.
\newblock {\em Monthly Notices of the Royal Astronomical Society},
  401(3):1933--1949, 2010.

\bibitem{burns_safronov}
J.A. {Burns} and V.S. {Safronov}.
\newblock {Asteroid nutation angles}.
\newblock {\em Monthly Notices of the Royal Astronomical Society}, 165:403--+,
  1973.

\bibitem{chang2014new}
Chan-Kao Chang, Adam Waszczak, Hsing-Wen Lin, Wing-Huen Ip, Thomas~A Prince,
  Shrinivas~R Kulkarni, Russ Laher, and Jason Surace.
\newblock A new large super-fast rotator:(335433) 2005 uw163.
\newblock {\em The Astrophysical Journal Letters}, 791(2):L35, 2014.

\bibitem{cicalo_YORP}
S~Cical{\`o} and DJ~Scheeres.
\newblock Averaged rotational dynamics of an asteroid in tumbling rotation
  under the yorp torque.
\newblock {\em Celestial Mechanics and Dynamical Astronomy}, 106(4):301--337,
  2010.

\bibitem{cotto_statler_YORP}
Desire{\'e} Cotto-Figueroa, Thomas~S Statler, Derek~C Richardson, and Paolo
  Tanga.
\newblock Coupled spin and shape evolution of small rubble-pile asteroids:
  Self-limitation of the yorp effect.
\newblock {\em The Astrophysical Journal}, 803(1):25, 2015.

\bibitem{gladman_2000}
B.~{Gladman}, P.~{Michel}, and C.~{Froeschl{\'e}}.
\newblock {The Near-Earth Object Population}.
\newblock {\em Icarus}, 146:176--189, July 2000.

\bibitem{goldreich}
P.~{Goldreich} and R.~{Sari}.
\newblock {Tidal Evolution of Rubble Piles}.
\newblock {\em Astrophysics Journal}, 691:54--60, January 2009.

\bibitem{golubov2014three}
O.~Golubov, D.J. Scheeres, and Y.N. Krugly.
\newblock {A Three-dimensional Model of Tangential YORP}.
\newblock {\em The Astrophysical Journal}, 794(1):22, 2014.

\bibitem{golubov_DSBYORP}
Oleksiy Golubov and Daniel~J Scheeres.
\newblock Equilibrium rotation states of doubly synchronous binary asteroids.
\newblock {\em The Astrophysical Journal Letters}, 833(2):L23, 2016.

\bibitem{harris_relax}
A.W. Harris.
\newblock Tumbling asteroids.
\newblock {\em Icarus}, 107(1):209--211, 1994.

\bibitem{hirabayashi_R3}
M.~{Hirabayashi}, D.~J. {Scheeres}, D.~P. {S{\'a}nchez}, and T.~{Gabriel}.
\newblock {Constraints on the Physical Properties of Main Belt Comet P/2013 R3
  from its Breakup Event}.
\newblock {\em ApJL}, 789:L12, July 2014.

\bibitem{toshi_1950DA}
M.~Hirabayashi and D.J. Scheeres.
\newblock {Stress and Failure Analysis of Rapidly Rotating Asteroid (29075)
  1950 DA}.
\newblock {\em The Astrophysical Journal Letters}, 798(1):L8, 2015.

\bibitem{holsapple_original}
KA~Holsapple.
\newblock Equilibrium configurations of solid cohesionless bodies.
\newblock {\em Icarus}, 154(2):432--448, 2001.

\bibitem{holsapple_plastic}
K.A. Holsapple.
\newblock Equilibrium figures of spinning bodies with self-gravity.
\newblock {\em Icarus}, 172(1):272--303, 2004.

\bibitem{holsapple_spinlimits}
K.A. {Holsapple}.
\newblock {Spin limits of Solar System bodies: From the small fast-rotators to
  2003 EL61}.
\newblock {\em Icarus}, 187:500--509, April 2007.

\bibitem{holsapple_YORP}
K.A. Holsapple.
\newblock {On YORP-induced spin deformations of asteroids}.
\newblock {\em Icarus}, 205(2):430--442, 2010.

\bibitem{jacobson_icarus}
S.~A. {Jacobson} and D.~J. {Scheeres}.
\newblock {Dynamics of rotationally fissioned asteroids: Source of observed
  small asteroid systems}.
\newblock {\em Icarus}, 214:161--178, July 2011.

\bibitem{jacobson_NEA_pop}
Seth~A Jacobson, Francesco Marzari, Alessandro Rossi, and Daniel~J Scheeres.
\newblock Matching asteroid population characteristics with a model constructed
  from the yorp-induced rotational fission hypothesis.
\newblock {\em Icarus}, 277:381--394, 2016.

\bibitem{jacobson_BYORP_equilibrium}
Seth~A Jacobson and D.J. Scheeres.
\newblock Long-term stable equilibria for synchronous binary asteroids.
\newblock {\em The Astrophysical Journal Letters}, 736(1):L19, 2011.

\bibitem{jacobson_mcmahon_wide}
Seth~A Jacobson, D.J. Scheeres, and Jay McMahon.
\newblock Formation of the wide asynchronous binary asteroid population.
\newblock {\em The Astrophysical Journal}, 780(1):60, 2014.

\bibitem{2017arXiv170309668J}
D.~{Jewitt}, J.~{Agarwal}, J.~{Li}, H.~{Weaver}, M.~{Mutchler}, and
  S.~{Larson}.
\newblock {Anatomy of an Asteroid Break-Up: The Case of P/2013 R3}.
\newblock {\em ArXiv e-prints}, March 2017.

\bibitem{kaasalainen_YORP}
Mikko Kaasalainen, Josef {\v{D}}urech, Brian~D Warner, Yurij~N Krugly, and
  Ninel~M Gaftonyuk.
\newblock Acceleration of the rotation of asteroid 1862 apollo by radiation
  torques.
\newblock {\em Nature}, 446(7134):420--422, 2007.

\bibitem{lowry_YORP}
Stephen~C Lowry, Alan Fitzsimmons, Petr Pravec, David Vokrouhlick{\`y}, Hermann
  Boehnhardt, Patrick~A Taylor, Jean-Luc Margot, Adrian Gal{\'a}d, Mike Irwin,
  Jonathan Irwin, et~al.
\newblock Direct detection of the asteroidal yorp effect.
\newblock {\em Science}, 316(5822):272--274, 2007.

\bibitem{paddack1969}
Stephen~J Paddack.
\newblock Rotational bursting of small celestial bodies: Effects of radiation
  pressure.
\newblock {\em Journal of Geophysical Research}, 74(17):4379--4381, 1969.

\bibitem{pravec2008spin}
P~Pravec, AW~Harris, D~Vokrouhlick{\`y}, BD~Warner, P~Ku{\v{s}}nir{\'a}k,
  K~Hornoch, DP~Pray, D~Higgins, J~Oey, A~Gal{\'a}d, et~al.
\newblock Spin rate distribution of small asteroids.
\newblock {\em Icarus}, 197(2):497--504, 2008.

\bibitem{pravec2014tumbling}
P~Pravec, P~Scheirich, J~{\v{D}}urech, J~Pollock, P~Ku{\v{s}}nir{\'a}k,
  K~Hornoch, A~Gal{\'a}d, D~Vokrouhlick{\`y}, AW~Harris, Emmanuel Jehin, et~al.
\newblock The tumbling spin state of (99942) apophis.
\newblock {\em Icarus}, 233:48--60, 2014.

\bibitem{pravec2016binary}
P~Pravec, P~Scheirich, P~Ku{\v{s}}nir{\'a}k, K~Hornoch, A~Gal{\'a}d, SP~Naidu,
  DP~Pray, J~Vil{\'a}gi, {\v{S}}~Gajdo{\v{s}}, L~Korno{\v{s}}, et~al.
\newblock Binary asteroid population. 3. secondary rotations and elongations.
\newblock {\em Icarus}, 267:267--295, 2016.

\bibitem{pravec_fission}
P.~Pravec, D.~Vokrouhlick{\`y}, D.~Polishook, D.J. Scheeres, AW~Harris,
  A.~Gal{\'a}d, O.~Vaduvescu, F.~Pozo, A.~Barr, P.~Longa, et~al.
\newblock Formation of asteroid pairs by rotational fission.
\newblock {\em Nature}, 466(7310):1085--1088, 2010.

\bibitem{pravec2002large}
Petr Pravec, Peter Kusnir{\'a}k, Lenka Sarounov{\'a}, Alan~W Harris, Richard~P
  Binzel, and Andrew~S Rivkin.
\newblock Large coherent asteroid 2001 oe84.
\newblock In {\em Asteroids, Comets, and Meteors: ACM 2002}, volume 500, pages
  743--745, 2002.

\bibitem{rossi_rotation}
A.~Rossi, F.~Marzari, and D.J. Scheeres.
\newblock {Computing the effects of YORP on the spin rate distribution of the
  NEO population}.
\newblock {\em Icarus}, 202(1):95--103, 2009.

\bibitem{rozitis2014cohesive}
B.~Rozitis, E.~MacLennan, and J.P. Emery.
\newblock {Cohesive forces prevent the rotational breakup of rubble-pile
  asteroid (29075) 1950 DA}.
\newblock {\em Nature}, 512(7513):174--176, 2014.

\bibitem{rubincam_YORP}
D.P. {Rubincam}.
\newblock {Radiative Spin-up and Spin-down of Small Asteroids}.
\newblock {\em Icarus}, 148:2--11, November 2000.

\bibitem{sanchez_icarus}
P.~S{\'a}nchez and D.J. Scheeres.
\newblock {DEM} simulation of rotation-induced reshaping and disruption of
  rubble-pile asteroids.
\newblock {\em Icarus}, 218:876--894, 2012.

\bibitem{sanchez_MAPS}
P.~S{\'a}nchez and D.J. Scheeres.
\newblock The strength of regolith and rubble pile asteroids.
\newblock {\em Meteoritics \& Planetary Science}, 49(5):788--811, 2014.

\bibitem{sanchez_disruption}
Paul S{\'a}nchez and Daniel~J Scheeres.
\newblock Disruption patterns of rotating self-gravitating aggregates: A survey
  on angle of friction and tensile strength.
\newblock {\em Icarus}, 271:453--471, 2016.

\bibitem{scheeres2017constraints}
D.~J. Scheeres.
\newblock Constraints on bounded motion and mutual escape for the full 3-body
  problem.
\newblock {\em Celestial Mechanics and Dynamical Astronomy}, 128(2):131--148,
  2017.

\bibitem{PSS_fission}
Daniel~J Scheeres.
\newblock Minimum energy asteroid reconfigurations and catastrophic
  disruptions.
\newblock {\em Planetary and Space Science}, 57(2):154--164, 2009.

\bibitem{scheeres_F2BP}
D.J. Scheeres.
\newblock Stability in the full two-body problem.
\newblock {\em Celestial Mechanics and Dynamical Astronomy}, 83(1):155--169,
  2002.

\bibitem{scheeres_fission}
D.J. Scheeres.
\newblock Rotational fission of contact binary asteroids.
\newblock {\em Icarus}, 189(2):370--385, 2007.

\bibitem{scheeres_YORP}
D.J. Scheeres.
\newblock {The dynamical evolution of uniformly rotating asteroids subject to
  YORP}.
\newblock {\em Icarus}, 188(2):430--450, 2007.

\bibitem{scheeres_itokawa_YORP}
D.J. Scheeres, M.~Abe, M.~Yoshikawa, R.~Nakamura, R.W. Gaskell, and P.A. Abell.
\newblock {The effect of YORP on Itokawa}.
\newblock {\em Icarus}, 188(2):425--429, 2007.

\bibitem{scheeres_AIV}
DJ~Scheeres, D~Britt, B~Carry, and KA~Holsapple.
\newblock Asteroid interiors and morphology.
\newblock {\em Asteroids IV}, pages 745--766, 2015.

\bibitem{scheeres_cohesion}
D.J. {Scheeres}, C.M. {Hartzell}, P.~{S{\'a}nchez}, and M.~{Swift}.
\newblock {Scaling forces to asteroid surfaces: The role of cohesion}.
\newblock {\em Icarus}, 210:968--984, December 2010.

\bibitem{scheirich_FG3}
P~Scheirich, P~Pravec, SA~Jacobson, J~{\v{D}}urech, P~Ku{\v{s}}nir{\'a}k,
  K~Hornoch, S~Mottola, M~Mommert, S~Hellmich, D~Pray, et~al.
\newblock The binary near-earth asteroid (175706) 1996 fg 3Ñan observational
  constraint on its orbital evolution.
\newblock {\em Icarus}, 245:56--63, 2015.

\bibitem{statler}
T.S. Statler.
\newblock {Extreme sensitivity of the YORP effect to small-scale topography}.
\newblock {\em Icarus}, 202(2):502--513, 2009.

\bibitem{taylor_margot_binaries}
Patrick~A Taylor and Jean-Luc Margot.
\newblock Binary asteroid systems: Tidal end states and estimates of material
  properties.
\newblock {\em Icarus}, 212(2):661--676, 2011.

\bibitem{taylor_YORP}
Patrick~A Taylor, Jean-Luc Margot, David Vokrouhlick{\`y}, Daniel~J Scheeres,
  Petr Pravec, Stephen~C Lowry, Alan Fitzsimmons, Michael~C Nolan, Steven~J
  Ostro, Lance~AM Benner, et~al.
\newblock Spin rate of asteroid (54509) 2000 ph5 increasing due to the yorp
  effect.
\newblock {\em Science}, 316(5822):274--277, 2007.

\bibitem{vokrouhlicky_tumbling}
D~Vokrouhlick{\`y}, S~Breiter, D~Nesvorn{\`y}, and WF~Bottke.
\newblock {Generalized YORP evolution: Onset of tumbling and new asymptotic
  states}.
\newblock {\em Icarus}, 191(2):636--650, 2007.

\bibitem{binary_AIV}
Kevin~J Walsh and Seth~A Jacobson.
\newblock Formation and evolution of binary asteroids.
\newblock {\em Asteroids IV. Univ. Arizona Press, Tucson}, pages 375--393,
  2015.

\bibitem{rotation_database}
B.D. Warner, A.W. Harris, and P.~Pravec.
\newblock The asteroid lightcurve database.
\newblock {\em Icarus}, 202(1):134--146, 2009.

\end{thebibliography}

\end{document}